# Thermodynamics of formation of multi-component CdZnTeSe solid solutions

S.V. Naydenov, I.M. Pritula

Institute for Single Crystals of the National Academy of Sciences of Ukraine,

60 Nauky Avenue, 61072 Kharkov, Ukraine

**Abstract**

A thermodynamic model of formation of multi-component solid solutions as a thermodynamic mixture of their binary components, is proposed. There are obtained expressions for the effective temperature of the equilibrium state of the solid solution of binary components, and for the excess Gibbs energy of formation of such a solution. The thermodynamic parameters of formation of solid solutions are calculated for $Cd_{1-x}Zn_xTe_{1-y}Se_y$ compounds with an arbitrary concentration of the doping elements. There is revealed a new thermodynamic effect of a decrease in the excess Gibbs energy in the solid solutions which contain "mixed" binary components formed during simultaneous doping of the cationic and anionic subsystems. It is shown that the excess Gibbs energy of formation of the quaternary CdZnTeSe solid solution (at additional doping with selenium) in comparing to the ternary CdZnTe compounds reaches the value –256 J/mol at the concentrations $x = 0.1$ and $y = 0.02$, that considerably exceeds the standard entropy of formation of pure binary components for the given solution. Such an effect is connected with emergence of strong covalent bonds in the "mixed" binary component ZnSe which is a part of the quaternary solid solution. This may explain a considerable decrease in the number of extended defects observed at the growth of CdZnTeSe crystals. The proposed theory allows different generalizations, and makes it possible to quantitively predict changes of the expected defect quality of the crystals at variations in the composition of the solid solutions.

1. Introduction

Crystals of different A2B6 compounds such as ZnSe, CdTe, CdSe, HgTe, etc. are widely used in the capacity of effective detectors of radiation in different spectrum ranges, see e.g. [1-7]. Ternary chalcogenide crystals, in particular, those of CdZnTe (CZT) solid solutions, were actively developed for uncooled room-temperature radiation detectors, see e.g. [1-4]. However, the quality of CZT crystals is limited, as they may contain a large number of extended defects [8]. Thereat, improvement of the crystal quality always gives rise to the detector efficiency [9]. As revealed in recent experiments, introduction of selenium into CZT crystals considerably reduces their defect structure [10]. Crystals of multi-component CdZnTeSe solid solutions are regarded as a viable alternative for up-to-date semiconductor detectors [11-15].

There arises the question concerning the physical causes of the fact that introduction of the additional solvent (selenium) into the multi-component CdZnTeSe compound leads to a strong improvement of the crystal quality. This issue has not yet been theoretically investigated. Proposed in the present study is a thermodynamic model that describes formation of solid solution of multi-component crystal (with the same valency of the elements in the chemical formula of the compound) as a "thermodynamic mixture" of all its possible binary components. Using CdZnTeSe as an example, there are calculated parameters of the model, in particular, the excess Gibbs energy of the solid solution formation. Comparison of the latter value shows an essential thermodynamic gain at a decrease in free energy for the quaternary CdZnTeSe crystals as against that for the ternary CdZnTe or CdTeSe crystals. The excess free energy accumulated during the obtaining of CdZnTe crystals at the transition to the equilibrium state will be relaxed by the system. Ordinarily, such a relaxation results in formation of numerous defects accompanied with minimization of the excess internal energy of the system. This essentially limits the quality of the obtained crystals.

Many macroscopic properties of crystals, peculiarities of their crystalline and defect structure phase transitions (state diagrams) depend on thermodynamic properties of the system, i.e. on thermodynamic functions and parameters. When going from simple compounds to complex ones, from one-component systems to complex multi-component systems, the thermodynamic properties of the latter can essentially vary. To a special extent this concerns liquid or solid solutions. At dissolution of even small amounts of some components in a certain solvent the properties of the formed solution may become radically different from those of its pure components. An eternal problem of thermodynamics and statistical physics is restoration of thermodynamic properties of molecular solution as a complex thermodynamic system, according to the parameters of their initial simple components [16]. Despite significant theoretical achievements in this direction (theories of ideal, slightly diluted, extremely diluted, regular and other types of solutions), full understanding of this problem is absent so far, especially when describing multi-component liquid solutions other than the simplest binary ones. For solid solutions attempts have been also made to build a complete theory. In particular, to describe solid solutions of three- and four- component A3B5 compounds [17] which are significant for semiconductor physics and technology. For multi-component solid solutions of A2B6 compounds such investigations have not been performed. As a rule, for many modern thermodynamic models there are applied computer methods of calculation based on the density functional method (or on its different variations). This requires the use of certain model interaction potentials which are very far from real interactions between atoms in complex non-molecular systems. In these calculations there are always introduced different fitting parameters which are initially absent in real experimental data. Thereat, clear understanding of physics of occurring phenomena requires building and application of self-consistent analytical (thermodynamic) models of multi-component solid solutions. This approach will make it possible to find out, restore and/or predict thermodynamic properties of solid solutions (at changing their composition) based on the set parameters of their simple, i.e. initially "pure" components. Unfortunately, thermodynamic models of such a type are still poorly developed.

The main problem at thermodynamic description of multi-component solid solutions is strong interaction between the crystal atoms with strong covalent and/or ionic bond. Direct determination of thermodynamic values based on calculation of the partition function or the density functional method does not guarantee a positive result. Thereat, the use of empirical models of ideal, weakly-ideal, slightly- and strongly-diluted solutions is not always possible. This is caused by the fact that at any solution concentration there always takes place a strong interaction between individual atoms of the solid solution (those of the matrix-solvent and the substituting atoms of the dissolved components). In the present work we restrict ourselves to studying the four-component solid solutions of $A_{1-x}X_xB_{1-y}Y_y$ type for A2B6 group compounds, due to significance of their application. In the capacity of initial matrix-solvent, the binary compound A2B6 is considered. The dissolved elements may be atoms of transient metals ($X^{2+}$ cations) or chalcogenide atoms ($Y^{2-}$ anions). Such compounds are $Cd_{1-x}Zn_xTe_{1-y}Se_y$ or $Cd_{1-x}Mn_xTe_{1-y}Se_y$ semiconductor crystals which are promising as new radiation sensitive or photosensitive materials. However, the proposed approach can be extended to other inorganic crystals. The concentration of dissolved components can vary over a fairly wide range – from fractions of a percent to tens of atomic percent. The solid solutions in question are initially characterized by strong covalent ionic bonds between all atoms of the crystal. Therefore, it is difficult to consider this solid solution as an ideal or slightly diluted solution containing several diluted atomic substitutional components. Each substitution atom in the crystal lattice is bonded to the nearest solvent atoms. Moreover, it can be also bonded to other substitutional atoms next to it, thereat the substituting cations are bonded to the substituting anions. Formation of such bonds can be interpreted as the existence and dissolution of certain binary components in the solid solution. For instance, for the system CdZnTeSe this signifies that in the main matrix-solution CdTe there are diluted three additional binary components – ZnTe (the substituting cation– the main anion of the solvent), CdSe (the main cation – the substituting anion), ZnSe (the substituting cation – the substituting anion). In general, within some idealization this solid solution can be considered ideal or slightly diluted solution which consists of binary components, since strong covalent bonds are already taken into account in the process of formation of the said binary components. Note that when going from the three-component solution to the four-component solution the thermodynamic configuration becomes essentially enriched, since the number of new independent thermodynamic components of the system significantly increases. Schematically this is shown in Fig. 1.

This approach is somewhat reminiscent of the fundamental concept of "quasiparticles" in physics of strongly interacting systems. The role of "quasiparticles" for a multi-component solid solution (as a system of strongly-interacting atoms) belongs to the binary components of this solution. Basically, the weak interaction between "quasiparticles", i.e. the binary solid solution components, can be taken into account within a suitable thermodynamic perturbation theory using some empirical approximations, i.e. expansions of thermodynamic values in powers of the concentration of the binary components, temperature and density. From the viewpoint of statistical physics, the transition of an atomic (non-molecular) solid solution to a solution of binary components makes it possible to reconsider it as almost "ideal gas" which consists of several weakly interacting binary (molecular) components. It is to be noted that the quasi-empirical calculation of thermodynamic values based on the data of the binary components has been earlier used for calculation of phase diagrams for three-component solid solutions, see e.g. [18, 19].

To verify the effectiveness of such an approach, it is necessary to build the corresponding thermodynamics and calculate the characteristic values that define the properties of the solid solution at changing its composition, based on the properties of its simple (binary) components. Thereat, there emerge interesting regularities. In particular, formation of solid solutions of three-component compounds does not result in any essential "thermodynamic gain" (decrease of the free energy of the system) in comparison with the one of the pure solvent. At the same time, in the solid solutions consisting of four or more components the said "thermodynamic gain" may take place. This means that the thermodynamic state of a four-component solid solution may be more energetically favourable, and, consequently, more stable and hardened as against the state of pure binary solvent or of its associated three-component solid solution. This effect has a direct physical consequence in the form of a reduction in the quantity of extended crystal defects which formation requires an excess free energy of the system. For instance, this takes place in CdZnTe crystal additionally doped with selenium. Apart from calculating the characteristic thermodynamic values (e.g. the excess Gibbs energy) which quantitatively describe the decrease in the free solid solution energy at dissolution of several binary components in it, it is important to establish their dependence on the composition, i.e. on the concentration of initial atomic components in the solid solution. This makes it possible to predict the expected change in the properties of multi-component solid solutions, and, consequently the quality of the grown crystals.

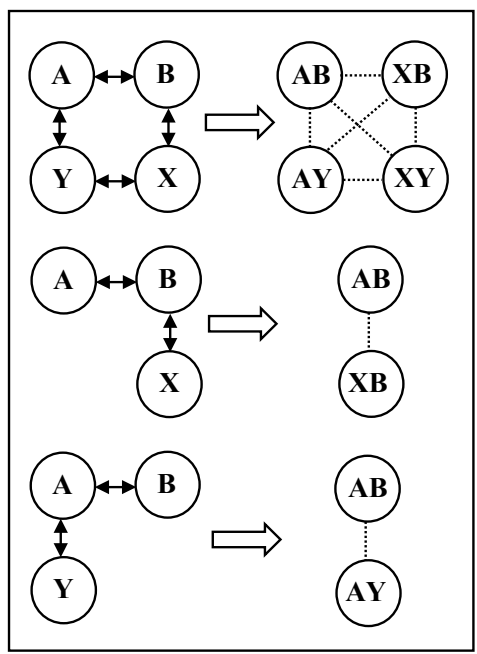

**Fig. 1.** Scheme of formation of solid solutions AXBY (four-component), AXB and ABY (three-component) which consist of several binary components. On the left: the initial atomic components with strong covalent bonds between them (arrows). On the right: the corresponding binary components with weak interaction (dotted lines)

## 2. Thermodynamic model of formation of multi-component solid solution

The key point of our thermodynamic approach is the idea to consider CdZnTeSe solid solution not traditionally, as a CdTe matrix containing dissolved simple mono-components (atoms) of Zn and Se, but as an ideal (or slightly diluted) solid solution obtained during dissolution of a number of binary components. Such a model is more valid from some viewpoint of physics. It allows to take into account the presence of strong covalent bonds between the doped Zn and Se atoms and the crystal lattice of main CdTe matrix. Because of which, it does not seem possible to consider CdZnTeSe as an ideal solution of CdTe diluted by a mixture of the mono-components of Zn and Se.

At rather high temperatures, formation of $Cd_{1-x}Zn_xTe_{1-y}Se_y$ compound from the melt can be formally considered to be the phase reaction

$$(1-\delta)CdTe + \alpha ZnTe + \beta CdSe + \gamma ZnSe \rightleftarrows Cd_{1-x}Zn_xTe_{1-y}Se_y \quad (1)$$

of dissolution of the three binary components ZnTe, CdSe and ZnSe in the solvent (melt) CdTe which is also a binary compound. In the state of thermodynamic equilibrium, the total Gibbs energy for thermodynamic mixture of binary components (the reaction reagents) is equal to the Gibbs energy of the solid solution (the reaction product). In a certain sense, the state of thermodynamic equilibrium is a formal abstraction of the proposed model. Moreover, the explicit state of "equilibrium of the binary components" may turn out to be physically difficult for implementation. Most of all, it corresponds to the concept of equilibrium between a solid solution (in the melt) and a mixture of saturated vapors of the corresponding binary compounds (above the melt).

Let us call each thermodynamic state of a solution or a melt with certain composition of binary components *binary state* of the system. Thereat, the concentration of the initial simple components Cd, Zn, Te, Se in the solution remains unchanged, and is set by a pair of fixed parameters $(x, y)$. At low primary concentrations $(x \ll 1, y \ll 1)$ a solution of binary components will also have a low concentration $(\alpha \ll 1, \beta \ll 1, \gamma \ll 1)$ of the dissolved components and a high concentration $1-\delta \approx 1$ of the solvent. In the first approximation it can be considered an ideal solution. At cooling of the melt, it goes into a normal solid solution.

For a stoichiometric compound there must be fulfilled the following relation

$$\begin{cases} x = \alpha + \gamma \\ y = \beta + \gamma \\ \delta = \alpha + \beta + \gamma \end{cases} \quad (2)$$

Note that the binary state of a solid solution depends not on the parameters $(x, y)$, but on the formal parameters $(\alpha, \beta, \gamma)$ with the bonding conditions imposed on them

$$\alpha - \beta = x - y \ . \tag{3}$$

Therefore, the thermodynamic potentials of the binary state ambiguously depend on the state parameters $(\alpha, \beta, \gamma)$. As usual, at thermodynamic equilibrium of the steady states are characterized by a conditional minimum of free energy of the system at variations of these parameters taking into account the additional relation (3).

Using the representation (1), the change in the free Gibbs energy when going from the initial binary components to the solid solution, i.e. the free energy of the solid solution formation can be written as

$$\Delta G_T(\alpha, \beta, \gamma) = \Delta H_T(\alpha, \beta, \gamma) - T \Delta S_T(\alpha, \beta, \gamma) \ , \tag{4}$$

where the change in the enthalpy and entropy of the system

$$\Delta H_T = \Delta_f H_T + \Delta_m H; \quad \Delta S_T = \Delta_f S_T + \Delta_m S \ , \tag{5}$$

is connected with the standard formation enthalpy $\Delta_f H^0$ and the standard formation entropy $\Delta_f S^0$ by the Kirchoff formulas [20]

$$\Delta_f H_T = \Delta_f H^0 + \int_{T_0}^{T} c_p(T) dT; \quad \Delta_f S_T = \Delta_f S^0 + \int_{T_0}^{T} c_p(T) \frac{dT}{T} \tag{6}$$

using the heat capacity $c_p$ of the components at constant pressure. The enthalpy $\Delta_m H$ and the melting entropy $\Delta_m S$ are connected by the equilibrium relation (see, for example [21])

$$\Delta_m H = T_m \Delta_m S \ , \tag{7}$$

where $T_m$ is the melting temperature of the substance.

The standard formation enthalpy and entropy for the solid solution components in binary state (in our terminology) are determined using the reactions

$$\begin{cases} Cd + Te \rightarrow CdTe \\ Zn + Te \rightarrow ZnTe \\ Cd + Se \rightarrow CdSe \\ Zn + Se \rightarrow ZnSe \end{cases} \tag{8}$$

and in accordance with the Hess law (see e.g. [22]), they are equal to

$$\begin{cases} \Delta_f H^0 = (1-\delta) \Delta_f H_*^0 + \alpha \Delta_f H_\alpha^0 + \beta \Delta_f H_\beta^0 + \gamma \Delta_f H_\gamma^0 \\ \Delta_f S^0 = (1-\delta) \Delta_f S_*^0 + \alpha \Delta_f S_\alpha^0 + \beta \Delta_f S_\beta^0 + \gamma \Delta_f S_\gamma^0 \end{cases}, \tag{9}$$

where $\Delta_f H_*^0$ and $\Delta_f S_*^0$ are the standard formation enthalpies and entropies for pure CdTe (solvent); as well as $\Delta_f H_\alpha^0, \Delta_f H_\beta^0, \Delta_f H_\gamma^0$ and $\Delta_f S_\alpha^0, \Delta_f S_\beta^0, \Delta_f S_\gamma^0$ are the standard formation enthalpies and entropies for pure ZnTe, CdSe, ZnSe binary components (solutes), correspondently. The melting enthalpy and entropy of solid solution of binary components are determined from similar formulas

$$\begin{cases} \Delta_m H = (1-\delta)\Delta_m H_* + \alpha\Delta_m H_\alpha + \beta\Delta_m H_\beta + \gamma\Delta_m H_\gamma \\ \Delta_m S = (1-\delta)\Delta_m S_* + \alpha\Delta_m S_\alpha + \beta\Delta_m S_\beta + \gamma\Delta_m S_\gamma \end{cases}, \quad (10)$$

where $\Delta_m H_i$ and $\Delta_m S_i$ are the melting enthalpies and entropies for the binary components numbered by the index $i = \{\alpha, \beta, \gamma\}$; $T_i$ are their melting temperatures and $\Delta_m H_i = T_i \Delta_m S_i$ in accordance with Eq. (7).

After certain transformations, the Gibbs energy (4) can be written as

$$\Delta G_T(\alpha, \beta, \gamma) = \Delta G_*(T) + [h_T(\alpha, \beta, \gamma) - s_T(\alpha, \beta, \gamma)T], \quad (11)$$

where values

$$\begin{cases} \Delta G_*(T) = \Delta_f G_*(T) + \Delta_m G_* \\ \Delta_f G_* = \Delta_f H_* - T\Delta_f S_* \\ \Delta_m G_* = \Delta_m H_* (1 - T/T_*) \end{cases} \quad (12)$$

correspond to the total change in the free Gibbs energy for the pure phase of CdTe solvent which includes the formation Gibbs energy $\Delta_f G_*$ of the solvent and the latent energy $\Delta_m G_*$ during its melting at temperature $T_*$. The equation (11) includes the auxiliary functions linearly depending (in the framework of our model of slightly diluted solid solution) on the partial concentrations,

$$\begin{cases} h_T(\alpha, \beta, \gamma) = \alpha(f_\alpha + g_\alpha) + \beta(f_\beta + g_\beta) + \gamma(f_\gamma + g_\gamma) \\ s_T(\alpha, \beta, \gamma) = \alpha(p_\alpha + q_\alpha) + \beta(p_\beta + q_\beta) + \gamma(p_\gamma + q_\gamma) \end{cases} \quad (13)$$

and the auxiliary parameters (constants at fixed temperature and pressure) numbered by the index $i = \{\alpha, \beta, \gamma\}$

$$\begin{cases} f_i = \Delta_f H_i - \Delta_f H_* \\ g_i = \Delta_m H_i - \Delta_m H_* \end{cases}; \quad \begin{cases} p_i = \Delta_f S_i - \Delta_f S_* \\ q_i = \dfrac{\Delta_m H_i}{T_i} - \dfrac{\Delta_m H_*}{T_*} \end{cases}. \quad (14)$$

It should be noted that in the general case the parameters $f_i, p_i$ (formation of binary components of a complex compound) depend on temperature, whereas the parameters $g_i, q_i$ (melting of binary components) are fixed.

Further we will not consider the dependence of the thermodynamic potentials on pressure, since for solid solutions under normal conditions this dependence is not strong. Moreover, to simplify the calculations we will assume that the value of heat capacity of the binary compounds is not temperature-dependent, i.e. $c_{p,i}(T) \approx c_{p,i}(T_0) = c_{p,i}^0$ at $T_0 \leq T \leq T_i$ up to the melting temperatures, where $T_0$ denotes the standard temperature. This approximation is rough enough, but in the present study it can be used for low concentrations $(x \ll 1, y \ll 1)$ of the solid solution. Then the corrections for the free energy due to the accepted approximation will have a higher order of smallness. Finally, let us assume that at $T \geq T_*$ above the melting temperatures of the solvent and of each of the binary components of the solid solution $T \geq T_i$, the formation latent heat (enthalpy) in the solution remains unchanged for those "pure" phases that have completely dissolved and disappeared in the melt. Though the enthalpy of the melt as a whole undoubtedly changes as temperature rises. Thereat, the formation entropy for the partial binary components contained in the melt is still essentially temperature-dependent. Thus, we assume that the mixing enthalpy for the binary components in the solution-melt is equal to zero (as it takes place for an ideal solution). Thereat, the mixing entropy for the solution-melt components changes with temperature. Such a fact resembles the model of *athermal solution* (see e.g. [16]). It remains valid under the assumption of ideal solution of binary components, if the concentration of the solved (melted) components remains low. There are no direct thermodynamic backgrounds that should be taken into account while choosing one of thermodynamic models for CdZnTeSe solution-melt. In particular, is to be considered a regular solution (the entropy of mixing the partial components in the melt is equal to zero), an athermal solution (the entropy of mixing these components in the melt is equal to zero), or an intermediate option? However, within our theory estimations for the model of regular solution leads to non-physical results, such as negative temperatures of the equilibrium state of the solution–melt consisting of binary components. So, there should be chosen the model of athermal solution. For correct definition of the thermodynamic type of CdZnTeSe solution-melt it is necessary to perform calorimetric measurements in the entire temperature range below and above the melting points of the binary components. Nowadays such data are absent.

From all approximations taking into account the relations (6) and the formulas (14), there follow approximate expressions for the introduced parameters, which in more general case depend on temperature:

$$\begin{cases} f_i \approx \left(\Delta_f H_i^0 - \Delta_f H_*^0\right) + \left(c_{p,i}^0(T_i - T_0) - c_{p,*}^0(T_* - T_0)\right) \\ p_i \approx \left(\Delta_f S_i^0 - \Delta_f S_*^0\right) + \left(c_{p,i}^0 - c_{p,*}^0\right)\ln\frac{T_i}{T_0} \end{cases}. \qquad (15)$$

Now they turn out to be quite definite with respect to the standard conditions. Here $c_{p,*}^0$ and $c_{p,i}^0$ denote the heat capacities of the solvent and the dissolved components under the standard conditions. When deriving the expressions (15) it is assumed that the melting temperature of the solvent substance is the lowest among those of all its binary components, i.e. we suppose that $T_* \leq T_i$ for all components. Under accepted conditions the earlier introduced functions $h_T, s_T$ become formally temperature-independent

$$h_T(\alpha,\beta,\gamma) \approx h(\alpha,\beta,\gamma); \quad s_T(\alpha,\beta,\gamma) \approx s(\alpha,\beta,\gamma) . \tag{16}$$

When solving the problem of conditional extremum for the potential (11) with the condition (3), we obtain the only independent relation, i.e. the equation for determination of the critical temperature $T^*$ of the equilibrium state of the system

$$T^* = \frac{(f_\alpha + g_\alpha) + (f_\beta + g_\beta) + (f_\gamma + g_\gamma)}{(p_\alpha + q_\alpha) + (p_\beta + q_\beta) + (p_\gamma + q_\gamma)} . \tag{17}$$

Let us call it the *effective temperature* of "equilibrium" solid solution (melt) of binary components. Strictly speaking, in the right-hand side of this equation there remains a weak temperature dependence for the parameters (14) which we neglect in certain calculations. In the general case, the equation (17) is the condition of thermodynamic self-consistency for a solid solution (melt) formed from binary components.

At $T = T^*$ the state of a solid solution (melt) formed by binary components should be considered as a new standard condition. Here the Gibbs energy of dissolution of binary components in such a solid solution acquires the value

$$\Delta G(T^*;\alpha,\beta,\gamma) = \Delta G_*(T^*) + \left[ h(\alpha,\beta,\gamma) - s(\alpha,\beta,\gamma)T^* \right] . \tag{18}$$

The excess Gibbs energy of solid solution as against that of pure solvent will be equal to

$$\Delta G^{ex}(T^*;\alpha,\beta,\gamma) \equiv \Delta G(T^*;\alpha,\beta,\gamma) - \Delta G_*(T^*) = h(\alpha,\beta,\gamma) - s(\alpha,\beta,\gamma)T^* , \tag{19}$$

For each type of solid solution with different (non-zero) parameters of the binary state $(\alpha,\beta,\gamma)$ the addition $\Delta G^{ex}$ will be determined at the same equilibrium temperature $T^*$. But for the compounds with different configuration of the components the temperature $T^*$ will differ. For instance, for the considered system

$$T^*(\text{CdZeTeSe}) = \frac{(f_\alpha + g_\alpha) + (f_\beta + g_\beta) + (f_\gamma + g_\gamma)}{(p_\alpha + q_\alpha) + (p_\beta + q_\beta) + (p_\gamma + q_\gamma)} , \tag{20}$$

but thereat

$$\begin{aligned} T^*(\text{CdZeTe}) &= \frac{(f_\alpha + g_\alpha)}{(p_\alpha + q_\alpha)}; \\ T^*(\text{CdTeSe}) &= \frac{(f_\beta + g_\beta)}{(p_\beta + q_\beta)}, \end{aligned} \tag{21}$$

whence it follows that in common case

$$T^*(\text{CdZeTeSe}) \neq T^*(\text{CdZeTe}) \neq T^*(\text{CdTeSe}) . \tag{22}$$

The excess Gibbs energy of two different types of solid solutions at the same temperature may essentially differ, even in the case when the content of the simple element that distinguishes these solutions is extremely small. For instance, in the case when $y \to 0$ or $x \to 0$ at transformation of $Cd_{1-x}Zn_xTe_{1-y}Se_y$ solid solution into $Cd_{1-x}Zn_xTe$ or $Cd_{1-x}Te_{1-y}Se_y$ solid solution, respectively.

The temperature $T^*$ corresponds to a certain formal state of thermodynamic equilibrium in the system of binary states. Therefore, it has no direct physical meaning, unlike the melting temperature of solid solutions or their pure binary components. As shown by the calculations (see below), $T^*$ for CdZnTeSe or CdZnTe solid solutions turns out to be higher than the maximum melting temperature of the binary components. At the same time, the excess Gibbs energy (19) already has the usual physical meaning of the change in the internal energy of the system during formation of a solid solution from partial (binary) components. Thereat it can be redefined using Kirchoff's formulas (similar to Eq. (6)) at any temperature, if we accept the hypothesis of the ideal solid solution formed from binary components. Therefore, within the proposed model we introduce the *excess Gibbs energy* of a solid solution formed from binary components

$$\Delta G_T^{ex}(\alpha,\beta,\gamma) \equiv h_T(\alpha,\beta,\gamma) - s_T(\alpha,\beta,\gamma)T \ , \qquad (23)$$

that bilinearly depends on the partial composition (with the condition $\alpha - \beta = x - y$) and temperature. Deviations of the solid solution from ideal model can be considered as the appearance of nonlinear terms (with respect to the concentrations and/or temperature) in this expression. Note that the definition of the excess energy (23) differs from the traditional one in the theory of thermodynamic solutions. It describes the change in the Gibbs energy in the process of dissolution of binary components in the solvent, but not the distinction of a real solution from an ideal one where the components are mixed in the solid or liquid phase, but not interact in the gas phase.

### 3. Dependence of excess Gibbs energy on binary composition of solid solutions

Let us consider the deviation of the introduced above excess Gibbs energy from its equilibrium value at $T = T^*$ depending on the solid solution composition. The corresponding expression (23) can be written in the form

$$\Delta G^{ex}(T^*;\alpha,\beta,\gamma) = \left[\alpha K_\alpha + \beta K_\beta + \gamma K_\gamma\right] , \qquad (24)$$

with the dimensional *constants of binary dissolution*

$$\begin{cases} K_\alpha = \dfrac{A_h(B_s + C_s) - A_s(B_h + C_h)}{A_s + B_s + C_s} \\ K_\beta = \dfrac{B_h(A_s + C_s) - B_s(A_h + C_h)}{A_s + B_s + C_s} \\ K_\gamma = \dfrac{C_h(A_s + B_s) - C_s(A_h + B_h)}{A_s + B_s + C_s} \end{cases}, \quad (25)$$

where new designations are introduced

$$\begin{cases} A_h = (f_\alpha + g_\alpha) \\ B_h = (f_\beta + g_\beta); \\ C_h = (f_\gamma + g_\gamma) \end{cases} \begin{cases} A_s = (p_\alpha + q_\alpha) \\ B_s = (p_\beta + q_\beta) \\ C_s = (p_\gamma + q_\gamma) \end{cases}. \quad (26)$$

For the dissolution constants there is fulfilled a significant linear relation

$$K_\alpha + K_\beta + K_\gamma = 0. \quad (27)$$

Using the relations (27) and (2) we obtain the following basic formula instead of the equation (24)

$$\Delta G^{ex}(x, y; \gamma) = -(xK_\beta + yK_\alpha) + 2\left(\frac{x+y}{2} - \gamma\right)(K_\alpha + K_\beta). \quad (28)$$

According to Eq. (2) parameter $\gamma$ may vary only within the limits $\min\{x, y\} \leq \gamma \leq \max\{x, y\}$. At the same time, at fixed $(x, y)$ the second term in the above expression always will have the fixed sign (positive or negative depending on the $\mathrm{sgn}(K_\alpha + K_\beta) = -\mathrm{sgn}\, K_\gamma$). The minimum of the free energy $\Delta G^{ex}$ at $(K_\alpha + K_\beta) > 0$ or the equivalent condition $K_\gamma < 0$ is reached at the highest concentration $\gamma = \gamma_{\max}$ of the "mixed" component in the solid solution. In the opposite case $K_\gamma > 0$ the minimum of $\Delta G^{ex}$ corresponds to the minimum concentration $\gamma = 0$ of this component.

The formula (28) describes the thermodynamic gain at the transition of the three-component solid solutions $Cd_{1-x}Zn_xTe$ or $CdTe_{1-y}Se_y$ to the four-component $Cd_{1-x}Zn_xTe_{1-y}Se_y$ solid solutions. In the former case the formula (28) does not contain negative "mixed" terms such as $\sim xK_\beta$ or $\sim yK_\alpha$. In this case values $K_\alpha, K_\beta$ should replace on the constants $\tilde{K}_\alpha, \tilde{K}_\beta$ of the extremely diluted solution obtained by removing one of the "mixed" binary components from it. As formally follows from the formula (25), $\tilde{K}_\alpha = \tilde{K}_\beta = 0$. As a result, within our model the excess Gibbs energy of a three-component solid solution formed by binary components turns out to be equal to zero

$$\Delta G^{ex}(\mathrm{CdZeTe}) = \Delta G^{ex}(\mathrm{CdTeSe}) = 0. \quad (29)$$

This is bound up with the used approximation of ideal solid solution formed by binary components. When considered separately, the components ZnTe or CdSe do not interact with the main component CdTe. In the four-component solution the components ZnTe, CdSe and ZnSe weakly interact with each other. The thermodynamic gain, i.e. the reduction of the free energy of the system $\Delta G^{ex} < 0$, may take place with an arbitrarily small addition of the "mixed" component ZnSe. As clearly seen from the formula (28), this depends on the ratio of the concentration of the initial components $x, y$ and the components $K_\alpha, K_\beta$. If $K_\alpha > 0, K_\beta > 0$ the gain will be the highest at $\gamma = (x+y)/2$, i.e. during formation of the maximum possible amount of the "mixed" component

$$\Delta G^{ex}_{opt}(x,y) \approx -(xK_\beta + yK_\alpha) . \qquad (30)$$

With simultaneous rise in the concentrations of both doping components $(x, y)$ in the solid solution such an effect will increase. It should be noted that the value $\Delta G^{ex}_{opt}$ of the revealed effect is a function of the thermodynamic state of the system, and depends solely on the nature and composition of the expected solid solution. It is independent of the type of the transition between nonequilibrium thermodynamic states of the system. That is why, e.g., at any method of the obtaining of solid solution crystals, there is a possibility to have a crystal with a minimum of the excess Gibbs energy. The degree of defectiveness of such a crystal will be the smallest.

The state in which the "mixed" binary component is not present, i.e. in the limit case $\gamma = 0$, is thermodynamically unstable at $K_\alpha > 0, K_\beta > 0$, since it is characterized by the positive excess Gibbs energy

$$\Delta G^{ex}(x, y; \gamma = 0) = (xK_\alpha + yK_\beta) . \qquad (31)$$

In the limiting case of $x \to 0$ or $y \to 0$, when the constants $K_\alpha, K_\beta$ tend to zero, there arises the state of neutral equilibrium with $\Delta G^{ex} \to 0$, which will be preserved for the corresponding three-component solid solution.

## 4. Thermodynamic comparing of formation of CdZnTe and CdZnTeSe solid solutions

Table 1 contains typical thermodynamic characteristics of the binary compounds (see e.g. [23, 24]) used in the capacity of binary components of CdZnTeSe solid solution. Based on these data, there were calculated thermodynamic parameters of our model for CdZnTeSe from the formulas (14)-(17) and (25)-(26). The obtained results are presented in Table 2 and Fig. 2. For comparison there are also given the parameters for the three-component solid solutions CdZnTe and CdTeSe. The effective "equilibrium" temperature $T^*$ of the solution consisting of the binary components was calculated from the formulas (20)-(22), for determination of the dimensional dissolution constants there were used the formulas (25). The negative sign of the dissolution constant points to the thermodynamic gain (decrease of the internal free energy of the system) at the formation of the corresponding binary component in the solution. The calculated values of the temperature $T^*$ of "equilibrium" state of the solution (melt) of binary components may turn out to be essentially higher. This is due to the fact that the approximations used for this estimation are rough enough. It concerns, in particular, the assumption that the heat capacity of the pure components is temperature independent.

It is to be noted that the effective temperature $T^*$ for the "equilibrium" solution (or, more precisely, melt) exceeds the melting temperature of all its binary components. This is quite natural from the viewpoint of the chosen approximation of the ideal solution consisting of binary components. The thermodynamic equilibrium between the pure phases (binary crystals in the melt) and the phase of solution (melt) with dissolved binary components can be obviously achieved when the melt temperature exceeds all the melting temperatures of the pure components. But this does not mean that complex crystals of solid solutions must be obtained at such high temperatures. The growth of the crystals from complex substances (synthesized by any convenient method) can be also performed at lower melting temperatures, as a rule, near the melting point $T_*$ of the pure solvent CdTe, or even at lower temperatures, in case of an excess of tellurium in the system. However, the thermodynamic effect bound up with the excess free energy $\Delta G^{ex}$ that arises during formation of the solid solution, will always be present, since it is due to the change in the thermodynamic state of the system, regardless of the way of its implementation. The case of $\Delta G^{ex} > 0$ testifies to thermodynamic instability and impossibility of formation of the solid solution that leads to phase separation. If $\Delta G^{ex} = 0$, the solution is extremely diluted with respect to one of its components and this corresponds to neutral state. The case of $\Delta G^{ex} < 0$ points to a gain in the internal energy and corresponds to the most stable thermodynamic state and to lower concentration of extended defects in the crystal.

**Table 1.** Thermodynamic values for binary A2B6 compounds (experiment)

| Parameter | CdTe | ZnTe | CdSe | ZnSe |
|---|---|---|---|---|
| Melting temperature, °C | 1042 | 1239 | 1263 | 1520 |
| $\Delta_f H^0$, 298 K, kJ/mol | -100.4 | -119.2 | -143.0 | -164.0 |
| $\Delta_f S^0$, 298 K, kJ/(mol·K) | 0.095 | 0.092 | 0.083 | 0.084 |
| $\Delta_m H$, kJ/mol | 44.0 | 65.0 | 44.0 | 67.0 |
| $c_p$, 298 K, kJ/(mol·K) | 0.0502 | 0.0497 | 0.0494 | 0.050 |

**Table 2.** Thermodynamic parameters of $Cd_{1-x}Zn_xTe_{1-y}Se_y$ solid solutions (calculations)

| Parameter | $Cd_{1-x}Zn_xTe$ | $CdTe_{1-y}Se_y$ | $Cd_{1-x}Zn_xTe_{1-y}Se_y$ |
|---|---|---|---|
| $T^*$, °C | 1735 | 1519 | 1636 |
| $K_A$, kJ/mol | 0 | 0 | 0.45 |
| $K_B$, kJ/mol | 0 | 0 | 2.47 |
| $K_C$, kJ/mol | 0 | 0 | -2.92 |
| $\Delta G^{ex}_{opt}$, kJ/mol | 0 | 0 | -0.256 (x=10%, y=2%) |

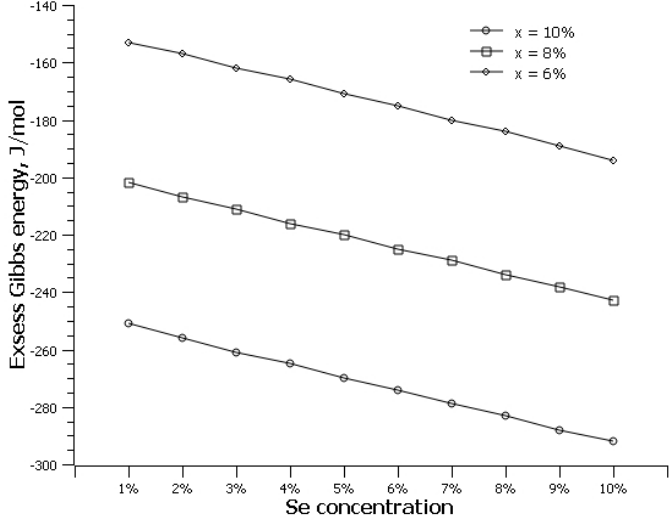

**Fig. 2**. The dependence of the excess Gibbs energy of $Cd_{1-x}Zn_xTe_{1-y}Se_y$ on binary composition.

The excess Gibbs energy $\Delta G^{ex}$ for CdZnTeSe solid solution with the concentrations of zinc $x = 10\%$ and selenium $y = 2\%$ was found to be equal to –256 J/mol. The negative sign points to a decrease in the internal free energy and to ordering of the system. The obtained value of $\Delta G^{ex}$ is comparable to the typical standard entropy values of formation of A2B6 binary compounds. Therefore, it testifies to an essential thermodynamic effect. The physical reason of large value of $\Delta G^{ex}$ is the presence of strong covalent bonds at formation of the "mixed" binary component ZnSe in the solid solution. In our opinion, it is just this effect that can give rise to a drastic reduction in defectiveness of CdZnTe crystals and in the number of their charge carrier traps at additional doping with selenium reported in [10-14]. In CdZnTe crystals the molar Gibbs energy is much higher by the value of the order of $\sim \left|\Delta G^{ex}_{CZTS}\right|$. During relaxation of the system to the equilibrium state this excess energy is redistributed to other types of the internal energy of the crystal. In particular, this leads to formation of dislocation networks and other extended defects which lower the overall free energy of the thermodynamic system.

Let us consider some peculiarities of the revealed effect of reduction of the excess Gibbs energy in the four-component CdZnTeSe solution obtained at additional doping of the three-component solid solution CdZnTe with selenium. The nature of the revealed effect is purely thermodynamic. Therefore, its consequences (a reduction in the number of possible defects, in particular, cadmium vacancies or tellurium inclusions/precipitates) are present in the crystals irrespective of the procedure of their obtaining, e.g. by High Pressure Bridgman Method (HPBM) or Heater Travelling Method (HTM) [25]. A reduction in the excess Gibbs energy takes place at any concentration of zinc and selenium in the solid solution, and increases as the concentration of zinc grows (see Fig. 2). This corresponds to the possibility to obtain high-quality four-component CdZnTeSe crystals with a very high (up to 20%) zinc concentration, even in the case when the concentration of selenium is negligible (near 2%). This cannot be achieved at the growth of the three-component crystals CdZnTe [26]. The excess Gibbs energy of the four-component solid solution monotonically reduces with increasing concentrations of zinc and selenium (see Eq. 30 and Fig. 2). This is caused by suppression of the defects both in the cation sublattice (vacancies of cadmium formed at substitution by zinc atoms) [25], and in the anion sublattice (precipitates of tellurium secondary phase which emerge as the solid solution is being formed) [10-12].

In early experiments it was found that large clusters (networks) of dislocations in CdZnTeSe crystals disappear for solid solutions with concentration of selenium higher than a certain threshold. As it follows from the previously obtained formula (28) the excess Gibbs energy vanishes $\Delta G^{ex}(x, y; \gamma_{cr}) = 0$ at the threshold concentration of "mixed" binary component

$$\gamma_{cr} = \frac{1}{2}(k_\alpha x + k_\beta y) , \qquad (32)$$

where the notations $k_\alpha = K_\alpha / (K_\alpha + K_\beta)$ and $k_\beta = K_\beta / (K_\alpha + K_\beta)$ are introduced. The relation $k_\alpha + k_\beta = 1$ and conditions $K_\alpha > 0, K_\beta > 0$ are also valid. We require that in an extremely dilute solutions at $x \ll 1$ or $y \ll 1$ there are two limits

$$\lim_{y \to 0} \gamma_{cr} = k_\alpha x; \quad \lim_{x \to 0} \gamma_{cr} = k_\beta y . \qquad (33)$$

These conditions are reminiscent of Henry's famous law (see, for example, [21]) for highly dilute solutions in thermodynamics. Physically, they determine the conditions of existence and thermodynamic stability of an extremely dilute solid solution, in which the concentration of one of the solvents is negligible. From the conditions (33) follows the formula

$$\left(\frac{x}{y}\right)_{cr} = \frac{k_\beta}{k_\alpha} = \frac{K_\beta}{K_\alpha} . \qquad (34)$$

It shows at what ratio of solvent concentrations in a dilute solid solution of binary components its excess Gibbs energy will be non-zero (in absolute value it will be minimal). Substitution of values from Table 1 gives $K_\beta / K_\alpha \approx 5.44$. Therefore, at given zinc concentration $x = 10\%$ in CdZnTeSe crystals we obtain the threshold concentration of selenium $y_{cr} \approx 1.8\%$ at which the previously described thermodynamic effect of strengthening of the solid solution (and reducing the number of extended defects) arises. For comparison, in recent experimental work [27], the threshold selenium concentration $y_{cr} \approx 1.5\%$ was proposed, starting from which many extended defects are practically absent in CdZnTeSe crystals.

As follows from the formula (30), with other conditions being equal, to decrease the free energy of the four-component crystal (and the quantity of possible defects in it), it is necessary to raise the concentration of selenium. However, in practice other factors should be taken into account, too. For instance, with the growth of the concentration of selenium the band gap of CdZnTeSe crystals decreases. This results in a rapid drop of the resistivity, which is unacceptable for high-resistance semiconductor radiation detectors. Therefore, there is some intermediate concentration of selenium which turns out to be optimal for the growth of detector quality crystals. For comparison note that, as show preliminary estimations, the excess Gibbs energy in CdMnTeSe solid solution containing manganese instead of zinc will have a nonmonotonic dependence on the mutual concentration of manganese and selenium. Consequently, expected manifestation of extended defects in such a system may be more complex, and the choice of the concentrations optimum for the crystal growth conditions may be less simple. Analysis of this system requires a special investigation.

## 5. Conclusions

Proposed is a thermodynamic model of formation of muti-component solutions of type of CdZnTeSe as a thermodynamic mixture of their binary components. The given model makes it possible to calculate some thermodynamic quantities for such a system, in particular, to determine the excessive Gibbs energy of formation of solid solution of binary components by referring to known thermodynamic characteristics of the partial binary compounds (the enthalpy and entropy of their formation and melting, as well as molar heat capacities). The value of the excess energy of mixing the binary components depends on the nature and composition of the solid solution. It is determined by the introduced constants of binary dissolution. Being a function of the thermodynamic state of the system, it is independent of the method that the solid solution has been obtained. Comparison of the said excess mixing energy for the quaternary CdZnTeSe solid solutions and the ternary CdZnTe, CdTeSe solid solutions reveals the thermodynamic effect of a severe free energy decrease in the quaternary solid solutions containing "mixed" binary component with co-doped cation (zinc) and anion (selenium) elements. The atomic nature of this effect lies in the formation of strong covalent bonds between zinc and selenium atoms homogeneously distributed in the solid solution which form the "mixed" binary component ZnSe dissolved in the main matrix of the solvent CdTe. Associated with this effect is the experimentally observed considerable improvement of the quality of CdZnTeSe crystals, namely, strong decrease in the defects imperfection of the crystals obtained under the optimum growth conditions.

The proposed approach allows generalization to various systems other than CdZnTeSe (in particular, to the systems CdMnTeSe with the addition of manganese instead of zinc), to those with other base (solvent), or with different number of the added cations (two or more), or with different valency of the elements, etc. Refinement of the model and more reliable calculations of the initial thermodynamic parameters of formation of the binary components in the solid solution, as well as the related dissolution constants necessitate realization of calorimetric measurements for the binary and multi-component compounds in a fairly wide temperature interval. Finally, the proposed model makes it possible to qualitatively and quantitatively predict changes in the thermodynamic properties of complex solid solutions, and consequently, the quality of the grown crystals depending on their composition.